\documentclass{paper} 
\usepackage{fullpage}
\usepackage{float}
\usepackage{booktabs}
\usepackage{multirow}
\usepackage{cite}

\usepackage{graphicx}
\usepackage[caption=false]{subfig}

\begin{document}

%
\title{Improving Automated COVID-19 Grading with Convolutional Neural Networks in Computed Tomography Scans: An Ablation Study}

\author{Coen~de~Vente,
        Luuk~H.~Boulogne \thanks{Coen~de~Vente and Luuk~H.~Boulogne contributed equally. (\textit{Corresponding author: L.~H.~Boulogne, e-mail: luuk.boulogne@radboudumc.nl})},
        Kiran~Vaidhya~Venkadesh,
        Cheryl~Sital,\\
        Nikolas~Lessmann,
        Colin Jacobs,
        Clara I. S\'{a}nchez,
        Bram~van~Ginneken \thanks{All authors are with the Radboud university medical center, Radboud Institute for Health Sciences, Department of Radiology and Nuclear Medicine, Nijmegen, The Netherlands.} 
}
\maketitle

\begin{abstract}
Amidst the ongoing pandemic, several studies have shown that COVID-19 classification and grading using computed tomography (CT) images can be automated with convolutional neural networks (CNNs). Many of these studies focused on reporting initial results of algorithms that were assembled from commonly used components. The choice of these components was often pragmatic rather than systematic. For instance, several studies used 2D CNNs even though these might not be optimal for handling 3D CT volumes.
This paper identifies a variety of components that increase the performance of CNN-based algorithms for COVID-19 grading from CT images. We investigated the effectiveness of using a 3D CNN instead of a 2D CNN, of using transfer learning to initialize the network, of providing automatically computed lesion maps as additional network input, and of predicting a continuous instead of a categorical output. A 3D CNN with these components achieved an area under the ROC curve (AUC) of 0.934 on our test set of 105 CT scans and an AUC of 0.923 on a publicly available set of 742 CT scans, a substantial improvement in comparison with a previously published 2D CNN. An ablation study demonstrated that in addition to using a 3D CNN instead of a 2D CNN transfer learning contributed the most and continuous output contributed the least to improving the model performance.

\end{abstract}

\begin{keywords}
deep learning, 3D convolutional neural network, COVID-19, CO-RADS, medical imaging.
\end{keywords}

\maketitle

\section{Introduction}
\begin{figure*}[!t]
	\centering
	\includegraphics[width=\linewidth]{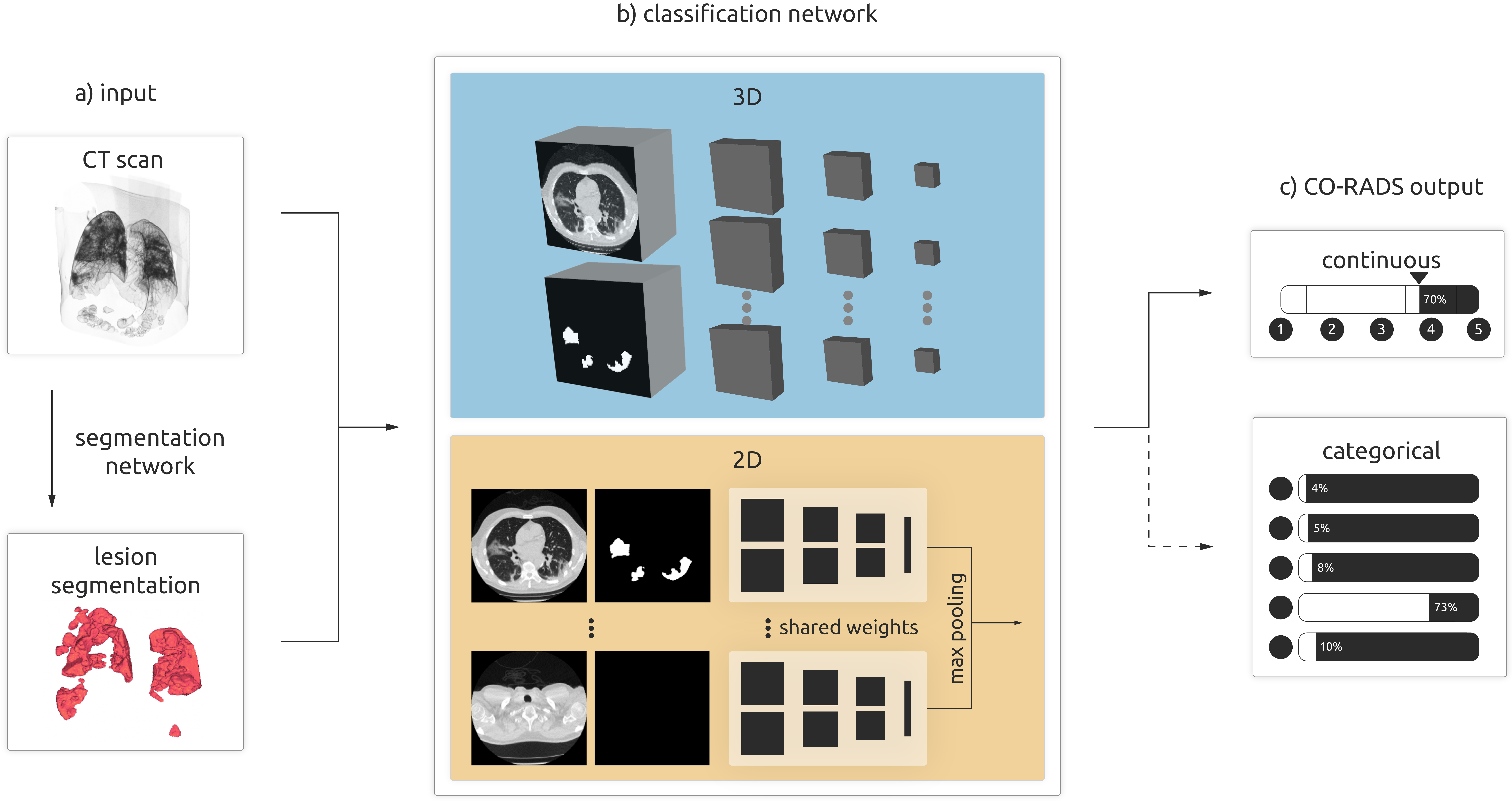}
	\caption{Summary of the processing pipeline. (a) The input CT scan is fed into a lesion segmentation network. The CT and the lesion segmentation are used as separate input channels to the classification network. In one of the ablation study experiments, we left out this lesion segmentation input. (b) We compared a 3D (top) with a 2D approach (bottom). The 3D network takes as input the full volume. The 2D network uses individual slices as input. (c) We compared a continuous output to a categorical output in the ablation study.}
	\label{fig:pipeline}
\end{figure*}

Imaging of COVID-19 with chest computed tomography (CT) has been found helpful for diagnosis of this disease in the current pandemic \cite{Yang2020}. With the aim to reduce the workload of radiologists, various machine learning techniques have been proposed to automatically grade and classify the presence of COVID-19 in CT images \cite{Li20,Butt20,Bars20,Wang20,Sing20a,Ying20,Wang20a,Jin20,Ouya20,Wang20b,Chen20,Zhen20a,Ning20,Less20,Sun20,Li20a}. Automatic COVID-19 classification methods have already been deployed in several medical centers \cite{Jin20}. 

The most common technique for automatic COVID-19 classification from CT images is the Convolutional Neural Network (CNN) \cite{Lecu90, Ji13a}, which is the current state-of-the-art for image classification \cite{Kriz12}. The works that use this approach can be divided into those that use 2D CNNs \cite{Li20,Butt20,Ying20,Wang20a,Chen20,Ning20} and those that use 3D CNNs \cite{Wang20,Ouya20,Wang20b,Zhen20a,Less20,Ning20,Li20a}. While 3D CNNs are directly capable of exploiting 3D information present in CT volumes, 2D CNNs can only indirectly use 3D information by aggregating the output of 2D networks for individual slices of the image to produce an image level prediction. 3D CNNs are typically more memory intensive than 2D CNNs, but Graphics Processing Units (GPUs) with sufficient memory to train 3D models are becoming increasingly available.
Moreover, radiologists are specifically instructed to take 3D information into account by inspecting different orthogonal views for assessing the suspicion of COVID-19 in CT scans \cite{Prok20}. This indicates that 3D information is essential for radiologists in assessing the patterns indicative for COVID-19. Additionally, the slice thickness of CT scans increasingly becomes smaller \cite{Ginn10a} so that the scans contain more detailed 3D information. We therefore hypothesize that 3D CNNs are more suitable for COVID-19 classification from CT scans than 2D CNNs.

In order to determine the optimal direction for the development of automatic COVID-19 classification systems, comparative studies need to be performed that evaluate different approaches. To indicate the generalization capabilities of automatic COVID-19 classification systems, some methods have been validated on data from different centers than the data that were used for training \cite{Wang20,Ouya20,Less20}. Also, the same validation methods, such as Receiver Operating Characteristic (ROC) curves and the area under the ROC curve (AUC), have been reported across different studies \cite{Li20,Wang20,Ying20,Wang20a,Jin20,Ouya20,Wang20b,Zhen20a,Less20,Ning20}. However, since each study used different datasets for training and for validation, a fair direct comparison of the performance of these algorithms is still not possible. However, the ``CT images and clinical features for COVID-19" (iCTCF) dataset was recently made publicly available and does allow for a fair comparison of COVID-19 classification methods \cite{Hust19}.

This paper compares a slice-based 2D CNN approach for COVID-19 classification to a 3D CNN approach. We trained and evaluated the approaches on the same internal dataset and additionally tested them on the iCTCF dataset to allow for a fair comparison between these algorithms. Moreover, we investigated performance changes due to 1) using transfer learning for 2D and 3D COVID-19 classification models, 2) using prior information in the form of COVID-19 related lesion segmentations as additional input to the network, 3) replacing the categorical output of our model with a continuous output. 

We furthermore created an open grand challenge \cite{Gccs20} for evaluating and comparing different COVID-19 classification algorithms. In this challenge, algorithms are evaluated on the iCTCF dataset that we used in this paper and can be compared to the methods presented in this paper, as well as to other COVID-19 grading and classification algorithms that are submitted to the challenge.

\section{Related Work}
3D CNNs were initially proposed for processing video data \cite{Ji13a}, where the third dimension of the convolutional layers dealt with the temporal dimension. In later work, 3D CNN architectures were derived from 2D CNN architectures by expanding the 2D filters into 3D \cite{Carr17}. Methods based on these inflated 3D CNNs, in particular the Inflated Inception V1 (I3D) model, have recently been successfully employed for lung nodule detection and scan-level classification tasks from thorax CT scans \cite{Ardi19,Hars20}.

Transfer learning (TL) is widely used in research on deep learning in medical imaging \cite{Ragh19a}. With TL, models are initialized with pre-trained weights from models trained on a different task or dataset. They are commonly pre-trained on the ImageNet \cite{Deng09} dataset that contains a large variety of 2D natural images. TL speeds up training and can offer performance gains for large models \cite{Ragh19a}. It has been used in several 2D CNN COVID-19 classification methods \cite{Li20,Ying20,Wang20a}. Pre-trained weights have also been used for 3D CNN-based methods. Wang \textit{et al.}~\cite{Wang20} applied TL by pre-training a model for COVID-19 classification on a large number of CT scans from lung cancer patients. I3D models can conveniently be initialized with inflated 2D weights. 2D weights have been used to pre-train I3D models for video classification \cite{Carr17} and chest CT classification \cite{Ardi19} tasks.

Before presenting CT images to the CNN, they are often pre-processed by extracting the lung region using lung or lobe segmentation algorithms. These lung regions are then used either for cropping around and centering to the lungs \cite{Less20,Wang20,Ying20,Li20a} and/or by suppressing non-lung tissue \cite{Li20,Butt20,Wang20,Ying20,Jin20,Ouya20,Wang20b,Zhen20a}. Lessmann \textit{et al.}~\cite{Less20} also added a lesion segmentation to the input of their model.

Previous works have trained models to discern between COVID-19 positive and negative patients \cite{Bars20,Sing20a,Ying20,Zhen20a}, COVID-19 positive patients and patients with other types of pneumonia \cite{Wang20,Wang20a,Ouya20}, and between all three \cite{Li20,Butt20,Wang20b}. In this work, we followed Lessmann \textit{et al.}~\cite{Less20} and trained our models to produce CO-RADS \cite{Prok20} scores on chest CT scans of suspected COVID-19 patients. The CO-RADS score denotes the suspicion of COVID-19 on a scale from 1 to 5 and was developed for standardizing reporting of CT scans of patients suspected with COVID-19 \cite{Prok20}. Scoring systems, like CO-RADS, have been advocated for better communication between radiologists and other healthcare providers \cite{Less20,Prok20}.

\section{Data}
\subsection{Training and internal test data}
The internal dataset contained CT scans from consecutive patients who presented at the emergency wards of the Radboud University Medical Center, the Netherlands in March, April and May 2020 and were referred for CT imaging because of moderate to severe COVID-19 suspicion. Medical ethics committee approval was obtained prior to the study. Further details such as imaging parameters can be found elsewhere \cite{Less20}.

CO-RADS scores were reported by a radiologist as part of routine interpretation of the scans. CO-RADS 1 was used for normal or non-infectious etiologies, having a very low level of suspicion. CO-RADS 2 was used if the CT-scan was typical for other infections than COVID-19, indicating a low level of COVID-19 suspicion. CO-RADS 3 implies equivocal findings and features compatible with COVID-19, but characteristics of other diseases are also found. CO-RADS 4 and 5 indicate a high and very high level of COVID-19 suspicion, respectively.

We randomly split the dataset into a development set with 616 patients and an internal test set of 105 patients. The patients in the development set were split into 75\% for training and 25\% for validation using data stratification based on the CO-RADS scores. The distribution of CO-RADS scores over the different splits is displayed in Table~\ref{tab:internaldataset}. All data splits were made such that all scans from a patient with multiple visits ended up in the same split.

\begin{table}[!t]
\centering
\caption{Number of CT Images in Internal Dataset.}
\label{tab:internaldataset}
\begin{tabular}{@{}llllllllll@{}}
\toprule
             &                        & \multicolumn{5}{c}{CO-RADS}                             &  &         \\ \cmidrule(lr){3-7}
             &                        & 1         & 2         & 3         & 4       & 5         & Total & Neg & Pos     \\ \midrule
\multicolumn{8}{l}{Development set}                                                                         \\
             & Training               & 253 & 71   & 78   & 37 & 73 & 512 & 324 & 188 \\
             & Validation             & 81   & 24  & 26   & 11 & 23 & 165 & 105 & 60  \\
\multicolumn{2}{l}{Internal test set} & 20   & 10  & 19   & 17 & 39 & 105 & 30  & 75  \\ \midrule
\multicolumn{2}{l}{Total}             & 354 & 105 & 123 & 65 & 135  & 782 & 459 & 323 \\
\bottomrule
\end{tabular}
\end{table}

\begin{table}[!t]
\centering
\caption{Number of CT Images in External Dataset.}
\label{tab:externaldataset}
\begin{tabular}{@{}llllllll@{}}
\toprule
                                    \multicolumn{5}{c}{Grade \cite{Ning20}}                             &  &         \\ \cmidrule(lr){1-5}
                                    Control         & Mild         & Regular         & Severe       & Critically ill         & Total & Neg & Pos     \\ \midrule
207 & 23   & 363   & 117 & 32 & 742 & 207 & 535 \\
\bottomrule
\end{tabular}
\end{table}

\subsection{External test data}
For external evaluation, we used the publicly available CT images and clinical features for COVID-19 dataset (iCTCF) dataset \cite{Hust19, Ning20}. Since we focused on comparing architectures for CT image processing for COVID-19 classification, we did not incorporate the clinical features from this dataset into the input for our models. In iCTCF, patients were categorized with a Chinese grading system that distinguishes the classes as Control, Mild, Regular, Severe, Critically ill and Suspected. Since there was no etiological evidence available for the presence of COVID-19 in Suspected cases \cite{Ning20}, we did not use them for testing our models. The distribution of the other classes is displayed in Table~\ref{tab:externaldataset}. The grading system uses etiological laboratory confirmation and other factors such as clinical features and CT imaging \cite{Ning20}. The control cases include both healthy patients and patients with community acquired pneumonia. Most of the iCTCF data has been made publicly available, but some CT scans were not available at the time of conducting this study. We validated our models with all available data from the first iCTCF cohort for which etiological evidence for the presence of COVID-19 was available \cite{Hust19}.



\section{Methods}

\subsection{2D and 3D architectures}
We compared 2D CNNs to 3D CNNs for the task of COVID-19 classification from CT scans as summarized in Fig.~\ref{fig:pipeline}. Since we used scan-level labels for training and testing our models, the 2D architecture required the integration of a slice-wise reduction step, while the 3D architecture did not. Therefore, we could not simply consider a 3D architecture and compare it with the same model in which the 3D convolutions were replaced with 2D convolutions. We instead compared two architectures that had been trained and evaluated before for the task of CT scan classification.

The 3D model we trained was 3D inflated Inception V1 (I3D) \cite{Carr17}, which was previously used as part of a system called CORADS-AI \cite{Less20}. 
The 2D model we trained was COVNet \cite{Li20}, which extracts features of individual axial slices using a 2D ResNet-50 architecture \cite{He16a}. A global max pooling step reduces these features to a 1D vector, to which a fully connected layer is applied with an output size equal to the number of classes. The code of COVNet has been made publicly available by the authors so that we were able to use the original implementation \cite{Covn20}.

\subsection{Lesion map as prior information}
To aid the model in localizing COVID-19 related parenchymal lesions, we provided a lesion segmentation map as additional input in an extra input channel. A 3D U-net which segments ground-glass opacities (GGOs) and consolidations \cite{Less20} provided these segmentations. GGOs and consolidations are biomarkers with major importance in diagnosing COVID-19 \cite{Prok20}.

\subsection{Pre-training}
For both the 2D and 3D CNNs, we used TL from natural image classification tasks. Following Lessmann \textit{et al.}~\cite{Less20}, the 3D model was initialized with a checkpoint from Carreira and Zisserman \cite{Carr17}, who first trained a 2D Inception V1 network on ImageNet \cite{Deng09} and afterwards inflated the model to 3D by replacing 2D convolution kernels with 3D kernels. Subsequently, this 3D model was further trained on RGB video data from the Kinetics dataset \cite{Kay17}. Following Li \textit{et al.}~\cite{Li20}, the ResNet-50 \cite{He16a} backbone of the 2D CNN that we applied to individual axial slices was also pre-trained on ImageNet.

\subsection{Continuous output}
The standard output format of CNNs used for categorical classification does not capture the ordinal nature of the CO-RADS scoring system. Furthermore, although the CO-RADS scoring system allows for a higher level of interpretability than a binary system, the fact that a CO-RADS suspicion score of 3 indicates that it is unclear whether COVID-19 is present makes it difficult to decide on the onset of the positive class for the predicted scores in ROC analyses. For these reasons, we considered the CO-RADS classification to be a regression task. Hence, the model had one output node that was forced to the range $(0,1)$ using the sigmoid function. CO-RADS scores were mapped to target values in the range $[0,1]$ with a uniform spacing between CO-RADS classes such that CO-RADS scores of 1 and 5 were assigned target values of 0 and 1, respectively. As the network had one output node, binary cross-entropy was used as loss function. With this method, unlike a standard categorical approach with a softmax layer and categorical cross-entropy loss, predictions that are further off from the target are penalized more heavily than predictions that are closer. To obtain a CO-RADS score during inference, the sigmoid output was multiplied by 4, rounded to the nearest integer and added to 1. De Vente \textit{et al.}~\cite{Vent20} explored this approach for prostate cancer grading and found that it outperformed other regression and categorical output methods.

\subsection{Pre-processing}
The CT scans were clipped between -1100 and 300 Hounsfield units, normalized between 0 and 1, and resampled to a voxel spacing of 1.5 mm$^3$ using linear interpolation. The scans were further pre-processed using a lung segmentation algorithm that was trained on data from patients with and without COVID-19 \cite{Xie20}. More specifically, any slices with a distance of 10 mm or more to the lung mask were discarded and the remaining slices were cropped to 240~$\times$~240 pixels around the center of the mask. Following previous research with I3D models \cite{Carr17,Ardi19,Hars20}, we trained our models with a fixed 3D input size. To achieve this without adding extra slices that do not contain information regarding the presence of COVID-19, we uniformly sampled 128 axial slices along the z-axis.

\subsection{Training}
We trained all networks with a batch size of 2, the Adam optimizer with $\beta_1$ = 0.9, $\beta_2$ = 0.999, and a learning rate of $10^{-4}$. Data augmentation consisted of zooming, rotation, shearing and elastic deformations in the axial plane, translation in all directions, and additive Gaussian noise. To correct for the class imbalance, we monitored the performance on the validation data in the development set during training with balanced samples based on the distribution of CO-RADS classes in the training set. We used early stopping with a patience of 10\,000 training batches and the quadratic weighted kappa (QWK) on the validation set for the stopping criterion. 
Gradient checkpointing \cite{Chen16e} was used to enable a batch size of 2 for the 2D model.

To rule out the possibility that performance differences between the 3D and 2D approach were due to other factors such as pre-processing or data augmentation, we only took the architecture implementation from the publicly available code for the 2D model \cite{Li20} and kept all other hyperparameters the same during training. Nevertheless, we also trained 2D models with the pipeline implemented by Li \textit{et al.}~\cite{Li20}. 
Unless otherwise specified, the results in Section~\ref{sec:results} were obtained with the prepossessing and training pipeline described here and not with the original pipeline.

Each model was trained on a single GPU, using NVIDIA GeForce GTX TITAN X, GeForce GTX 1080, GeForce GTX 1080 Ti, GeForce RTX 2080 Ti, and TITAN Xp cards.

\subsection{Ensembling}
The models were sensitive to the randomness of the training process introduced by initialization of weights without pre-training, sample selection, and data augmentation. In order to enable stable comparisons, we obtained ensembles by training 10 instances of the same model with different random seeds. The ensemble output was obtained by taking the mean of the individual model outputs. For categorical model ensembles, the output was the mean of the probability output vectors of the individual models.
All results presented in Section~\ref{sec:results} were obtained from ensembles unless stated otherwise.

\subsection{Evaluation}
We evaluated the CO-RADS scoring performance using the QWK score. This measure accounts for the ordinal nature of the CO-RADS score by weighting mismatches between true and predicted labels differently based on the magnitude of the error. Following previous works on COVID-19 classification and grading \cite{Li20,Wang20,Ying20,Wang20a,Jin20,Ouya20,Wang20b,Zhen20a,Less20,Ning20}, diagnostic performance was evaluated using the AUC and ROC curves. 

We calculated 95\% confidence intervals (CIs) with non-parametric bootstrapping and $1000$ iterations \cite{Rutt00}. Statistical significance was computed with the same bootstrapping method \cite{Samu07a}. 

The AUCs that our models achieved on the external test set are additionally listed on the Grand Challenge platform \cite{Gccs20} to allow for a direct comparison between our and future COVID-19 grading and classification solutions.

    
    

\section{Results}
\label{sec:results}

\begin{figure}[!b]
	\centering
	\includegraphics[width=\linewidth]{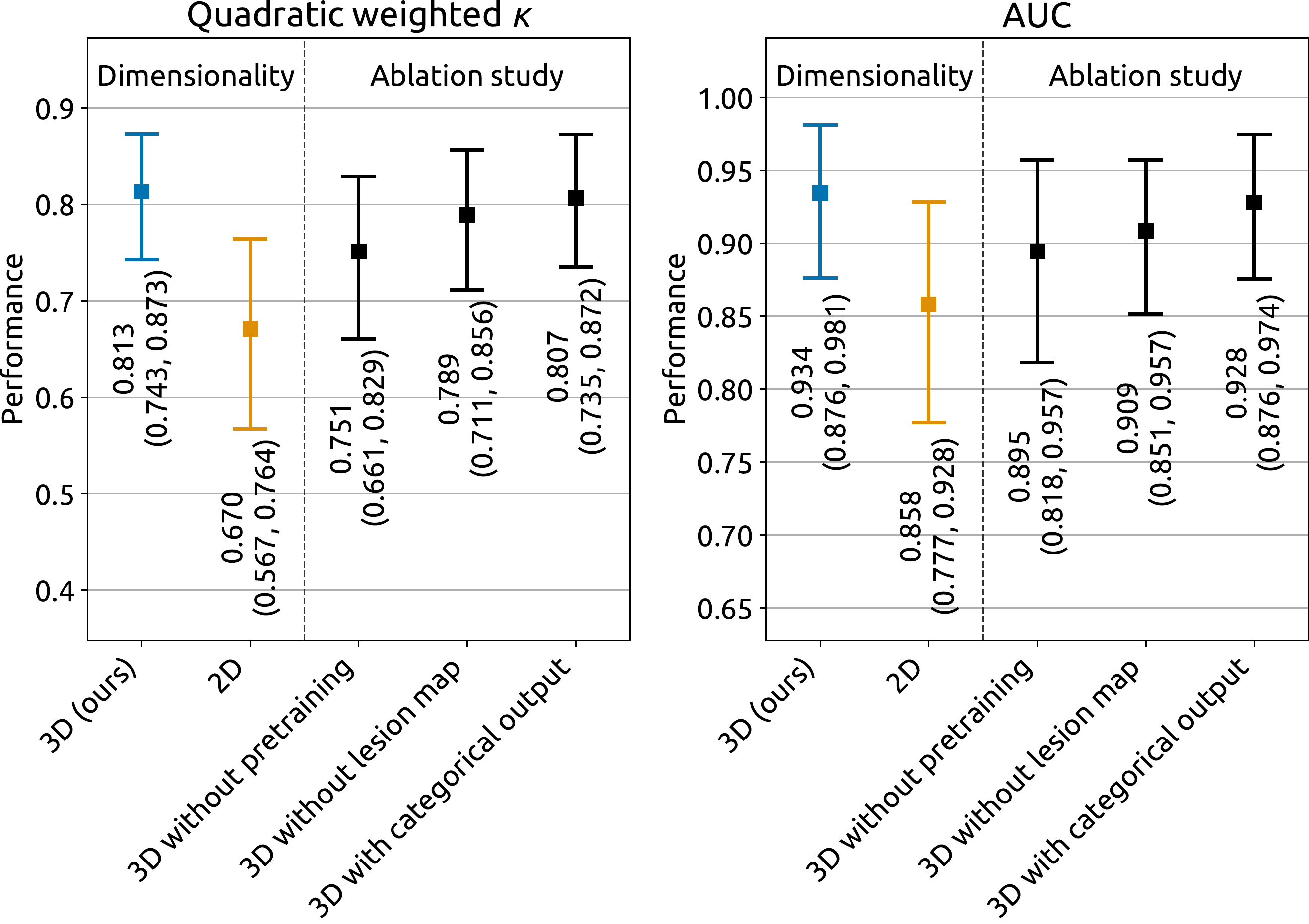}
	\caption{Comparison of 2D and 3D CNNs and ablation study for automatic COVID-19 grading from CT images. The analysis was performed on the internal test set. The error bars indicate the 95\% CIs. The AUC was computed with CO-RADS 1-2 as the negative class (30 scans) and CO-RADS 3-5 as the positive class (75 scans).}
	\label{fig:ablation}
\end{figure}

\begin{figure*}[!t]
    \centering
    \subfloat[]{
        \includegraphics[width=\linewidth]{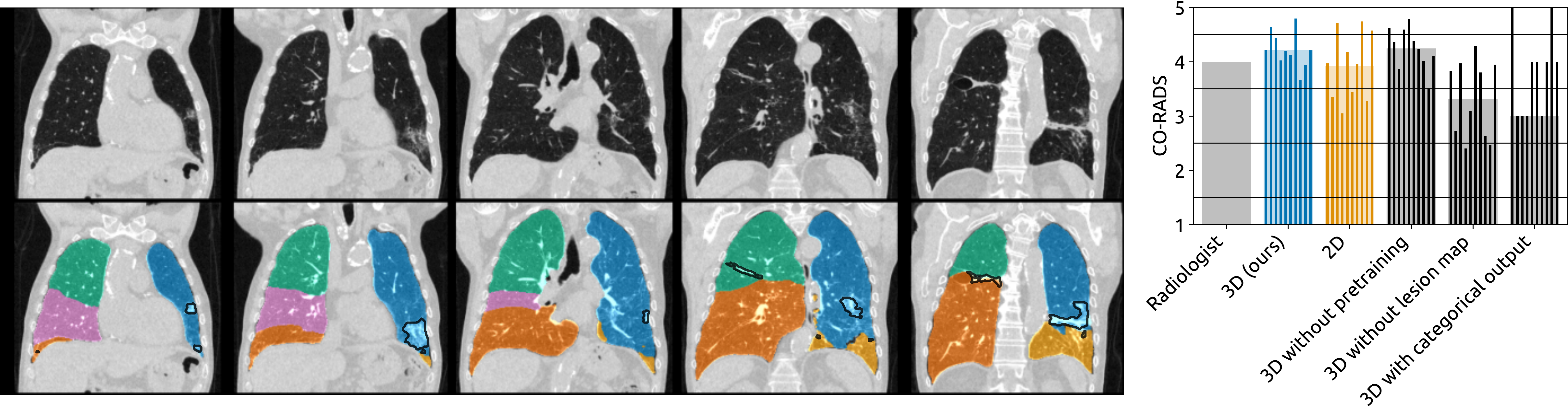}
        \label{fig:10099_st000}
        }
    \\
    \subfloat[]{
        \includegraphics[width=\linewidth]{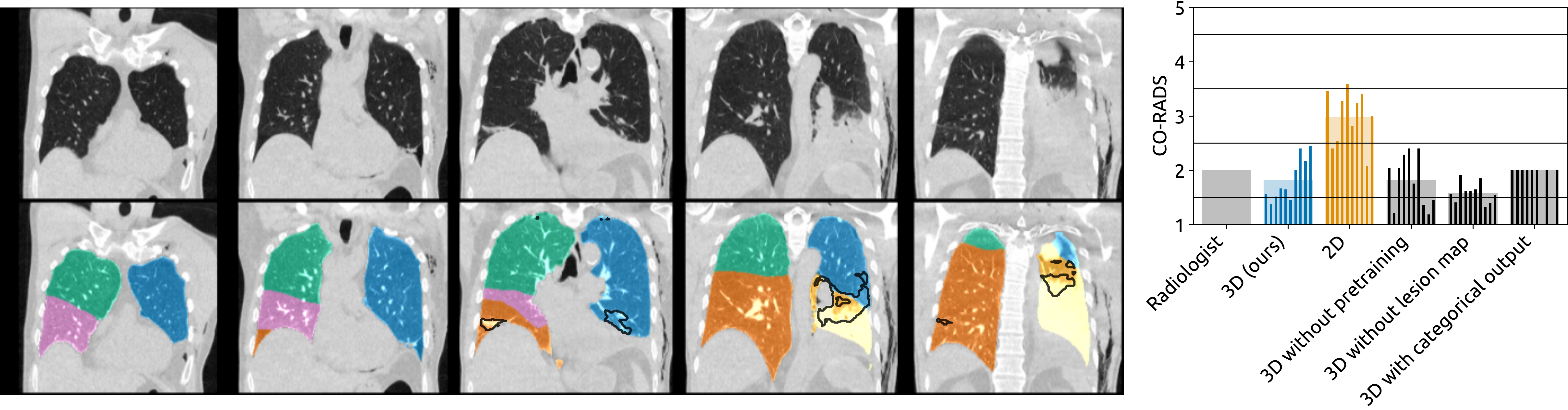}
        \label{fig:10005_st000}
        }
    \\
    \subfloat[]{
        \includegraphics[width=\linewidth]{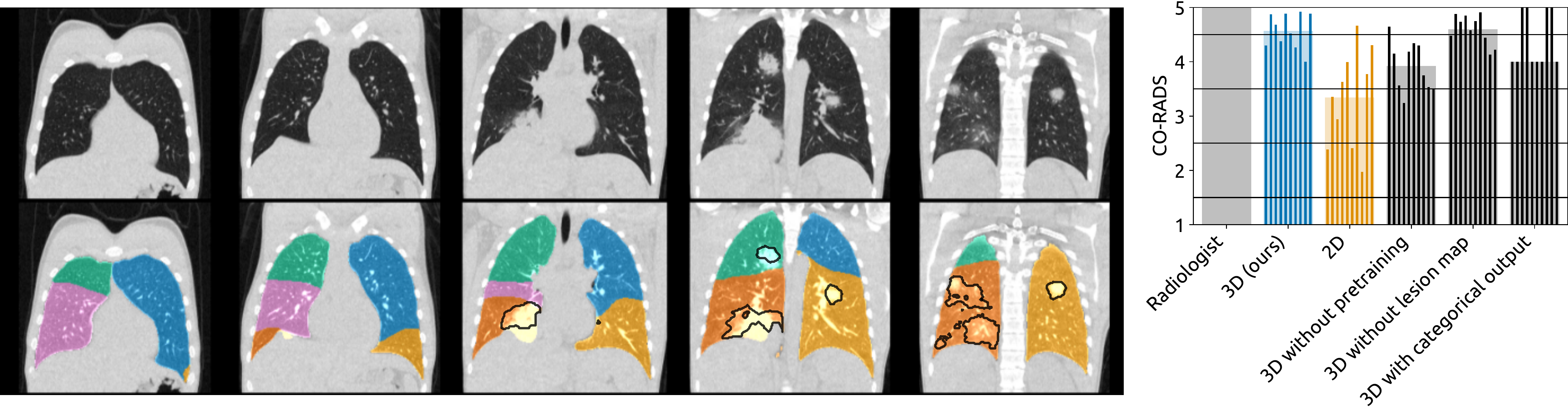}
        \label{fig:10071_st001}
        }
    \caption{Example input-output pairs of the internal test set for the trained ensembles. Input examples are shown on the left. Top row: Coronal slices of an input CT scan. Bottom row: Lung segmentation used for centering and cropping are displayed with colored overlays. Delineations of the lesion masks that were used as a separate input channel are depicted as black lines. Output examples of the ensembles (wide, light bars) as well as the individual models these ensembles are composed of (narrow, dark bars) are shown on the right. (a) Radiology report: "GGO and consolidations especially lower lobes and posterior. Has had prior lung carcinoma. COVID-19 is probable, but other infection intrapulmonal is also possible." (b) Radiology report: "COVID-19 not probable, but also not ruled out. Known post-traumatic thorax, persistent pleura fluid, slice pneumothorax. Small amount of GGO and consolidation (left). Some pneumonia at thorax trauma, post-traumatic deviations." (c) Consolidation and GGO in all lobes. According to radiologist: "Very suggestive for COVID. Also positive PCR. Proven comorbidity." }
    \label{fig:input_output_examples}
\end{figure*}

\subsection{2D vs. 3D CNNs}
\begin{figure}[!t]
	\centering
	\includegraphics[width=.8\linewidth]{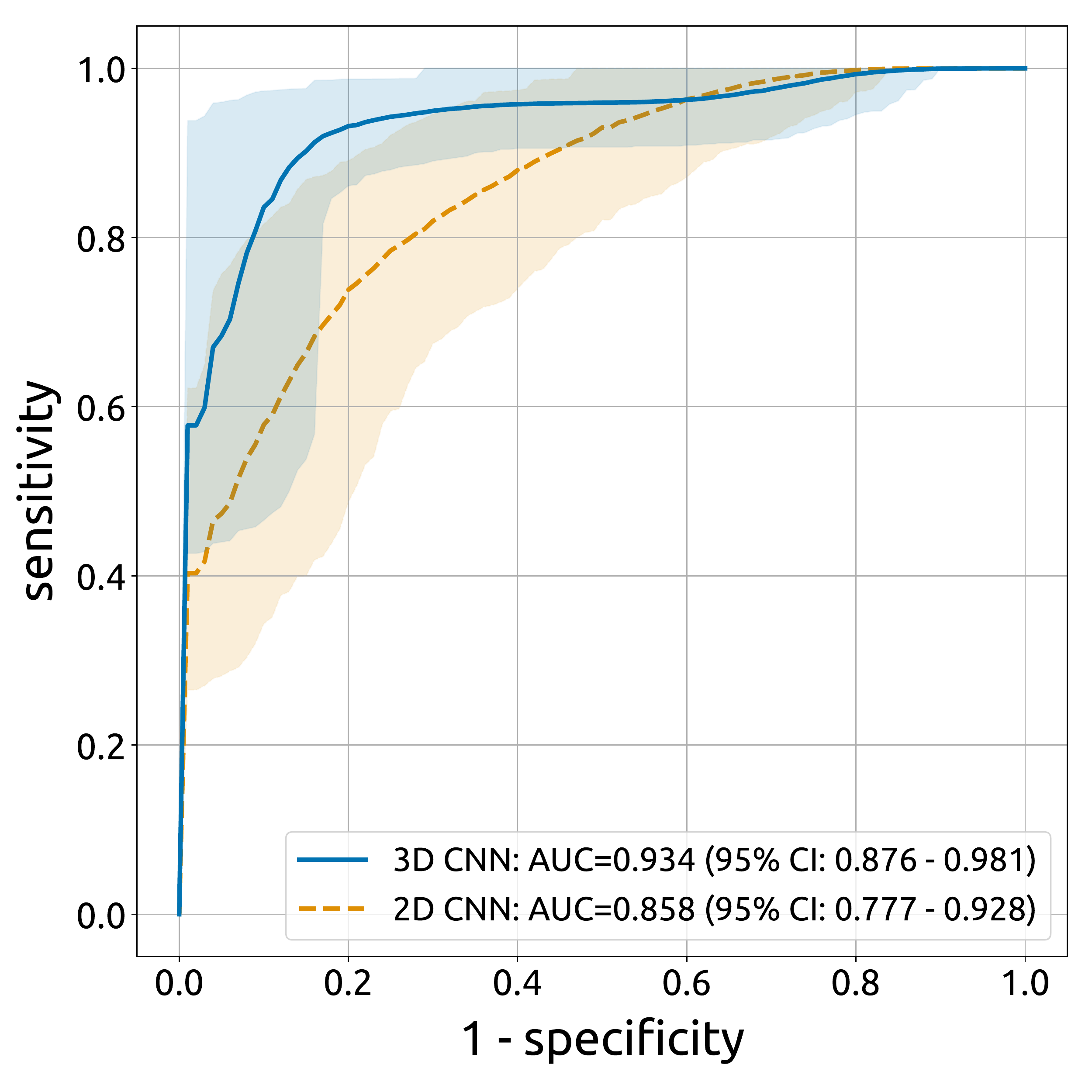}
	\caption{ROC analysis for the internal test set from Radboudumc (105 CT scans). The analysis was performed  with CO-RADS 1 and 2 as the negative class (30 scans) and CO-RADS 3-5 as the positive class (75 scans). It was performed for the full 2D and 3D models trained with transfer learning, lesion maps and continuous output.}
	\label{fig:roc_internal}
\end{figure}

On the internal dataset, both the AUC and the QWK scores were significantly higher for the full 3D model (with transfer learning, lesion maps and continuous output) than for the full 2D model ($p<.001$ for both metrics). Figures \ref{fig:ablation} and \ref{fig:roc_internal} show the corresponding CIs and ROC analyses respectively. Fig.~\ref{fig:input_output_examples} shows prediction examples from the full 3D, full 2D and ablated 3D models in blue, yellow, and black respectively.

When training an ensemble with the pipeline from Li \textit{et al.}~\cite{Li20} we obtained a lower performance on the internal test set than when we applied the 2D model in our own pipeline. With the pipeline from Li \textit{et al.}, we obtained a QWK of 0.567 (95\% CI: 0.411-0.703, $p=.054$) and a lower AUC of 0.828 (95\% CI: 0.741-0.906, $p=.206$).

\begin{figure}[!b]
    \centering
    \includegraphics[width=\linewidth]{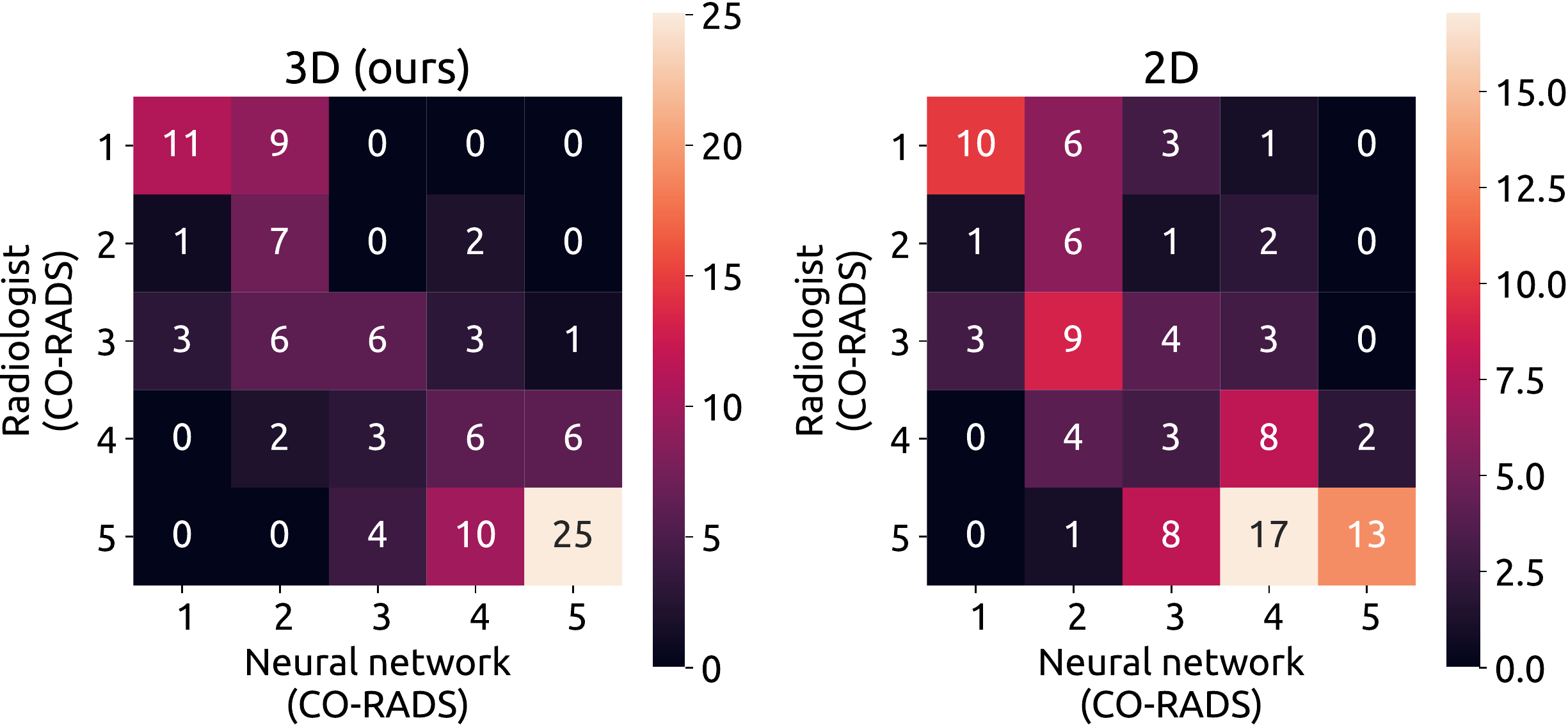}
    \caption{Confusion matrices for CO-RADS scoring of the 2D and 3D model predictions on the internal test set. These models were trained with transfer learning, lesion maps and produced continuous output. The true label reference is from the radiology report. Cells contain the number of CT scans.}
	\label{fig:confusion_internal}
\end{figure}

Fig.~\ref{fig:confusion_internal} shows confusion matrices for the two dimensionalities. For 13 scans, the full 3D approach had predictions that were more than one CO-RADS category off. For the full 2D approach this was the case for 24 scans. Furthermore, the full 3D approach had no cases that were further off than 2 categories, while the full 2D approach did have two of these cases.

On average, training of the individual 3D models took approximately 16 hours (21\,100 iterations), while it took about 30 hours (30\,000 iterations) for the 2D models.


\begin{figure}[!t]
	\centering
	\includegraphics[width=.8\linewidth]{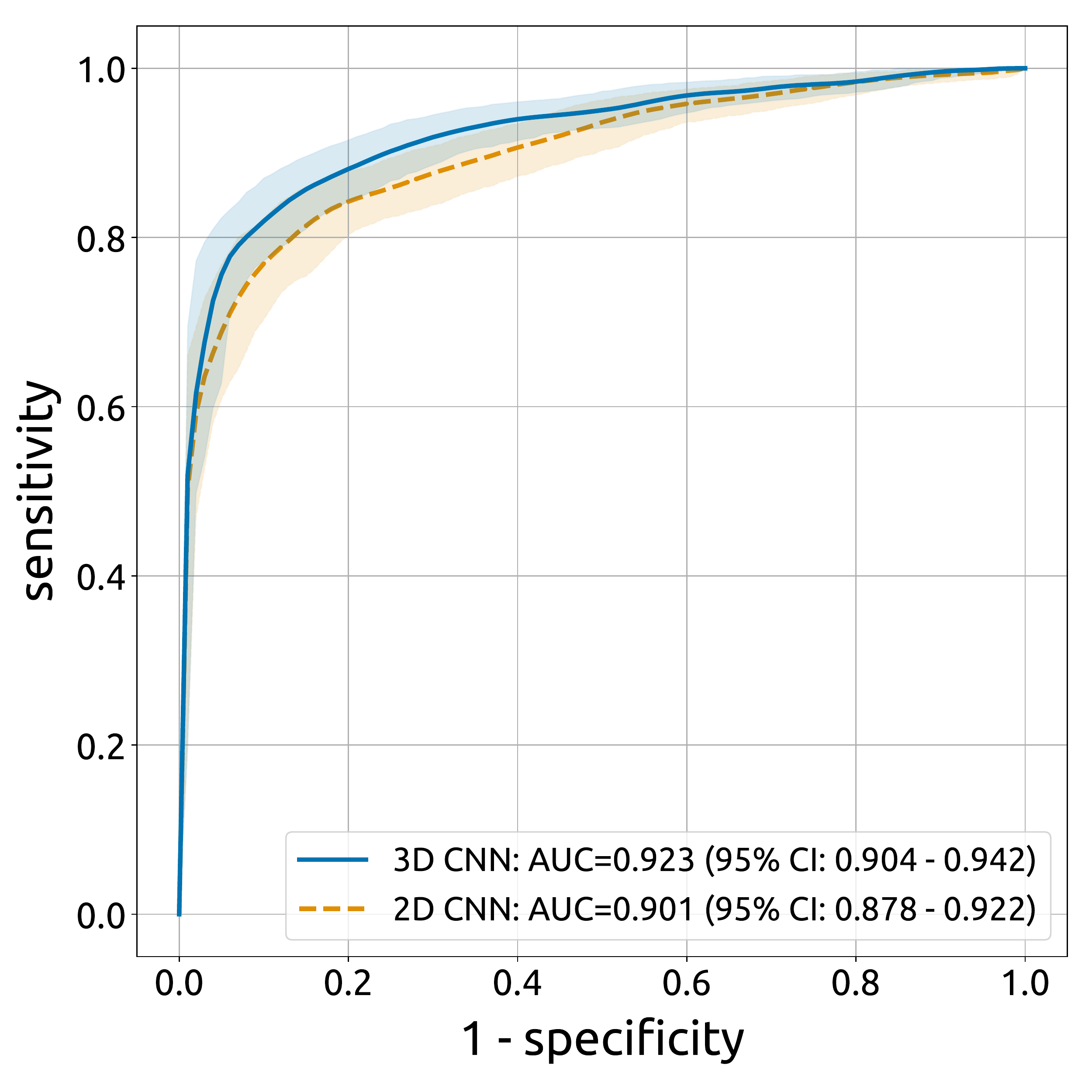}
	\caption{ROC analysis for the external iCTCF test set (742 CT scans). The analysis was performed with 207 COVID-19 negative (Control) cases and 535 positive (Mild, Regular, Severe, Critically ill) cases.}
	\label{fig:roc_external}
\end{figure}

\subsection{Ablation study}
The results of an ablation study to investigate the effect of each of the additional components added to the 3D CNN are shown in Fig.~\ref{fig:ablation}. Removing any of the additions had a negative effect on both the AUC and the QWK, which were 0.934 and 0.813 for the 3D model without ablations respectively. The biggest drop in performance occurred when removing pre-training from the full model, which reduced the AUC to 0.895 ($p=.004$) and the QWK to 0.751 ($p=.011$). Removing the lesion segmentation input from it reduced the AUC to 0.909 ($p=.116$) and the QWK to 0.789 ($p=.220$). Replacing the regression approach with a categorical target had the smallest effect on performance, reducing the AUC to 0.928 ($p=.237$) and the QWK to 0.807 ($p=.397$). Fig.~\ref{fig:input_output_examples} shows prediction examples from the ablation study models in black. The models without pre-training and without lesion input needed more iterations than the 3D model, which required 21\,100 iterations on average. Their mean numbers of iterations were 26\,050 and 26\,300, respectively. The categorical output, however, only needed 12\,950 iterations to finish training on average.

\subsection{External validation}
Fig.~\ref{fig:roc_external} shows the ROC curves of the full 3D and the full 2D model for the external iCTCF test set. The 3D approach outperformed the 2D approach in terms of AUC ($p = .001$). An ablation experiment confirmed that the full 3D model performed better than the models without pre-training, without lesion map input, and with categorical output, which achieved AUCs of 0.923 (95\% CI: 0.901 - 0.942, $p=.440$), 0.920 (95\% CI: 0.898 - 0.940, $p=.343$), and 0.910 (95\% CI: 0.889 - 0.930, $p=.0140$), respectively.



\section{Discussion}
In this paper, we identified and tested components of CNN based automated COVID-19 grading models. More specifically, we investigated how the performance of such a model is affected by using 3D CNNs instead of 2D CNNs, using transfer learning, using automatically computed lesion maps as additional network input, and predicting a continuous output instead of a categorical output. We evaluated all models with the same datasets to allow for a fair comparison between models. 
 
The full 3D model (with transfer learning, lesion maps and continuous output) outperformed the full 2D model in terms of AUC and QWK score on the internal test set for COVID-19 classification and CO-RADS grading.

For the 2D architecture we used COVNet, an architecture previously used in a similar COVID-19 classification task in CT \cite{Li20}, for which the authors reported an AUC of 0.96 for differentiating between COVID-19 positive and negative patients. The substantial difference between this result and our observations with COVNet, both when using only their architecture (AUC = 0.858) and when using both their architecture and their full training pipeline (AUC = 0.828), illustrates the importance of using the same dataset when comparing different approaches.

The 3D model with categorical output obtained an AUC of 0.928 (95\% CI: 0.876 - 0.974) on the internal test set. This model is similar to the CO-RADS scoring component of CORADS-AI \cite{Less20}, which obtained an AUC of 0.95 on the same test set. This AUC is slightly higher, but lies within the 95\% CI stated above. However, Lessmann \textit{et al.}~\cite{Less20} computed AUC using only the CO-RADS 5 probability, while in our work we used CO-RADS 1 and 2 as the negative class and CO-RADS 3-5 as the positive class. Furthermore, Lessmann \textit{et al.} did not use ensembling and used a slightly different pre-processing method. Therefore, this may not be a fair comparison.

We also observed a better diagnostic performance for COVID-19 classification by the 3D model on the external test set, where the AUC was 0.923 for the full 3D model, while it was 0.901 for the full 2D model. Ning \textit{et al.}~\cite{Ning20} developed a 2D model with slice-level annotations indicating if the slice was COVID-19 positive, negative or non-informative. Using a superset of the external set used in this paper for evaluation an AUC of 0.919 was obtained, which is lower than the AUC of our proposed 3D model. This further underlines the importance of using 3D rather than 2D models.

A possible explanation for why adding the extra dimension to the convolutions improves the performance is that it allows the CNN to take into account the 3D structure and full volume of individual lesions. This explanation is in line with the fact that radiologists typically use both the axial and coronal views to visualize the spread of COVID-19 related lesions across the lungs in CT scans, such as GGOs \cite{Prok20}.

We could not directly compare the CO-RADS classification performance on the external set, since CO-RADS labels were not available. Moreover, the CO-RADS grading cannot be directly translated to the system used in the iCTCF dataset, since the former measures the probability of COVID-19 presence, while the latter quantifies the severity of the disease.

The internal test set was comprised of data from the same population as the data the model was trained on, while the external test set was comprised of data from a different population. For the full 2D model, a lower AUC was obtained on the internal test set than on the external test set. This difference might be due to population differences between the internal and external test set, or due to the different definitions of the positive class, which were presence of COVID-19 and high suspicion of COVID-19 for the internal and external test sets respectively. 

On the external test set, the full 3D model outperformed the full 2D model by a smaller margin in terms of AUC than on the internal dataset. This difference could be partly due to the different definitions of the positive class. However, we also found that it partly arises from the larger overall slice thickness in the external test set. All scans in the internal test set had a slice thickness of 0.5mm. In contrast, 207 scans (40 COVID-19 positive, 167 negative scans) in the external test set had a slice thickness larger than 1.5mm, which was the input resolution in our training and testing pipeline. When evaluating only on these scans, we obtained an AUC of 0.886 (95\% CI: 0.836-0.929) for the full 3D model and an AUC of 0.873 (95\% CI: 0.819-0.922) for the full 2D model. The external test set contained 535 scans (167 COVID-19 positive, 368 negative) with a slice thickness smaller than or equal to 1.5mm. On these scans we obtained an AUC of 0.931 (95\% CI: 0.907-0.951) for the full 3D model and an AUC of 0.905 (95\% CI: 0.879-0.929) for the full 2D model. The performance of both models is lower for scans with a large slice thickness, but this effect is more apparent for the 3D model.

The ablation study on the internal test set showed that our proposed additions to the network and training procedure have a positive effect on the performance. On the external test set, the ablation study showed only small improvements in AUC due to these additions. These performance increases were smaller than the performance increases obtained for these components on the internal test set. As with the difference in performance of the 2D model on the internal and external test set, these differences could be due to population differences between the internal and external test set and the different definitions of the positive class in these datasets.

Although the ablation studies for the internal and external test sets both showed an increase in performance for each component, not all performance increases were statistically significant. For a significance level of 0.05, the increase in QWK was insignificant for both adding the lesion segmentation input and replacing the categorical output with the continuous output. The increase in AUC was insignificant for replacing the categorical output with the continuous output on the internal test set, for adding pre-training on the external test set, and for adding lesion maps as input for both test sets. Except for using lesion maps as input, all components thus resulted in a significant performance increase for at least one of the test sets. Regardless of performance increases, using a continuous output removes the disadvantage of  having to decide on the onset of the positive class for the predicted CO-RADS scores. Adding lesion maps as input might be ineffective for automated CNN based COVID-19 grading methods.

Our results show that TL from natural 2D images and video data is beneficial for COVID-19 classification. However, other pre-training schemes can be explored in future work. For example, employing TL by pre-training with another dataset with large amounts of thorax CT scans as was done in \cite{Wang20} might lead to a further increase in performance.

We did not use clinical features available for the external dataset as input to the models trained in this work, since the main goal of this paper was to demonstrate the effect on performance of different COVID19 grading and classification algorithm components. In future work, however, clinical data could be combined with our proposed 3D approach.

%
\section{Conclusion}
We compared 2D and 3D Convolutional Neural Network (CNN) architectures for COVID-19 classification from computed tomography scans and found that the 3D architecture outperforms the 2D architecture on an internal and an external dataset. We investigated how the performance of our model was affected by including COVID-19 related lesion segmentations as additional input, using transfer learning, and replacing the categorical output of our model with a scalar continuous output. 

The models and the automatic evaluation method we used in this paper have been made available on the online Grand Challenge platform \cite{Gccs20}. This allows researchers to obtain and compare the performance of their COVID-19 grading and classification solutions to other solutions on the platform.

Our findings advance the performance of automated COVID-19 grading systems and provide insight in the performance benefits of several of their components. These insights primarily indicate that future research and clinical applications should move towards using 3D CNNs for COVID-19 grading in CT scans.

\bibliographystyle{abbrv}
\bibliography{fullstrings,diag,diagnoweb,newrefs}

\begin{thebibliography}{10}

\bibitem{Hust19}
{CT} images and clinical features for {COVID}-19.
\newblock {http://ictcf.biocuckoo.cn/HUST-19.php} Accessed: 2020-05-20.

\bibitem{Covn20}
{GitHub} - bkong999/{COVNet}: Artificial intelligence distinguishes {COVID}-19
  from community acquired pneumonia on chest {CT}.
\newblock {https://github.com/bkong999/COVNet} Accessed: 2020-08-24.

\bibitem{Gccs20}
{Grand Challenge} - {COVID-19 CT Classification} challenge.
\newblock {https://covid19.grand-challenge.org/} Accessed: 2020-09-01.

\bibitem{Ardi19}
D.~Ardila, A.~P. Kiraly, S.~Bharadwaj, B.~Choi, J.~J. Reicher, L.~Peng, D.~Tse,
  M.~Etemadi, W.~Ye, G.~Corrado, D.~P. Naidich, and S.~Shetty.
\newblock End-to-end lung cancer screening with three-dimensional deep learning
  on low-dose chest computed tomography.
\newblock {\em Nature medicine}, 25:954--961, 2019.

\bibitem{Bars20}
M.~Barstugan, U.~Ozkaya, and S.~Ozturk.
\newblock Coronavirus ({COVID}-19) classification using {CT} images by machine
  learning methods.
\newblock {\em arXiv:2003.09424}, 2020.

\bibitem{Butt20}
C.~Butt, J.~Gill, D.~Chun, and B.~A. Babu.
\newblock Deep learning system to screen coronavirus disease 2019 pneumonia.
\newblock {\em Applied Intelligence}, page~1, 2020.

\bibitem{Carr17}
J.~Carreira and A.~Zisserman.
\newblock Quo vadis, action recognition? a new model and the {Kinetics
  Dataset}.
\newblock In {\em Proceedings of the IEEE Conference on Computer Vision and
  Pattern Recognition}, pages 6299--6308, 2017.

\bibitem{Chen20}
J.~Chen, L.~Wu, J.~Zhang, L.~Zhang, D.~Gong, Y.~Zhao, S.~Hu, Y.~Wang, X.~Hu,
  B.~Zheng, et~al.
\newblock Deep learning-based model for detecting 2019 novel coronavirus
  pneumonia on high-resolution computed tomography: a prospective study.
\newblock {\em medRxiv}, 2020.

\bibitem{Chen16e}
T.~Chen, B.~Xu, C.~Zhang, and C.~Guestrin.
\newblock Training deep nets with sublinear memory cost.
\newblock {\em arXiv:1604.06174}, 2016.

\bibitem{Vent20}
C.~de~Vente, P.~Vos, M.~Hosseinzadeh, J.~Pluim, and M.~Veta.
\newblock Deep learning regression for prostate cancer detection and grading in
  bi-parametric {MRI}.
\newblock {\em IEEE Transactions on Biomedical Engineering}, 2020.

\bibitem{Deng09}
J.~Deng, W.~Dong, R.~Socher, L.~Li, K.~Li, and L.~Fei-Fei.
\newblock Imagenet: A large-scale hierarchical image database.
\newblock In {\em Computer Vision and Pattern Recognition}, pages 248--255,
  2009.

\bibitem{Hars20}
I.~W. Harsono, S.~Liawatimena, and T.~W. Cenggoro.
\newblock Lung nodule detection and classification from thorax {CT}-scan using
  {RetinaNet} with transfer learning.
\newblock {\em Journal of King Saud University-Computer and Information
  Sciences}, 2020.

\bibitem{He16a}
K.~He, X.~Zhang, S.~Ren, and J.~Sun.
\newblock Deep residual learning for image recognition.
\newblock In {\em Proceedings of the IEEE conference on computer vision and
  pattern recognition}, pages 770--778, 2016.

\bibitem{Ji13a}
S.~Ji, W.~Xu, M.~Yang, and K.~Yu.
\newblock 3d convolutional neural networks for human action recognition.
\newblock {\em IEEE Transactions on Pattern Analysis and Machine Intelligence},
  35:221--231, 2013.

\bibitem{Jin20}
S.~Jin, B.~Wang, H.~Xu, C.~Luo, L.~Wei, W.~Zhao, X.~Hou, W.~Ma, Z.~Xu,
  Z.~Zheng, et~al.
\newblock Ai-assisted ct imaging analysis for {COVID}-19 screening: Building
  and deploying a medical {AI} system in four weeks.
\newblock {\em medRxiv}, 2020.

\bibitem{Kay17}
W.~Kay, J.~Carreira, K.~Simonyan, B.~Zhang, C.~Hillier, S.~Vijayanarasimhan,
  F.~Viola, T.~Green, T.~Back, P.~Natsev, et~al.
\newblock The {Kinetics} human action video dataset.
\newblock {\em arXiv:1705.06950}, 2017.

\bibitem{Kriz12}
A.~Krizhevsky, I.~Sutskever, and G.~Hinton.
\newblock Imagenet classification with deep convolutional neural networks.
\newblock In {\em Advances in Neural Information Processing Systems 25}, pages
  1097--1105, 2012.

\bibitem{Lecu90}
Y.~LeCun, B.~E. Boser, J.~S. Denker, D.~Henderson, R.~E. Howard, W.~E. Hubbard,
  and L.~D. Jackel.
\newblock Handwritten digit recognition with a back-propagation network.
\newblock In {\em Advances in neural information processing systems}, pages
  396--404, 1990.

\bibitem{Less20}
N.~Lessmann, C.~I. Sanchez, L.~Beenen, L.~H. Boulogne, M.~Brink, E.~Calli,
  J.-P. Charbonnier, T.~Dofferhoff, W.~M. van Everdingen, P.~K. Gerke,
  B.~Geurts, H.~A. Gietema, M.~Groeneveld, L.~van Harten, N.~Hendrix,
  W.~Hendrix, H.~J. Huisman, I.~Isgum, C.~Jacobs, R.~Kluge, M.~Kok,
  J.~Krdzalic, B.~Lassen-Schmidt, K.~van Leeuwen, J.~Meakin, M.~Overkamp,
  T.~van Rees~Vellinga, E.~M. van Rikxoort, R.~Samperna, C.~Schaefer-Prokop,
  S.~Schalekamp, E.~T. Scholten, C.~Sital, L.~St\"{o}ger, J.~Teuwen,
  K.~Vaidhya~Venkadesh, C.~de~Vente, M.~Vermaat, W.~Xie, B.~de~Wilde,
  M.~Prokop, and B.~van Ginneken.
\newblock Automated assessment of co-rads and chest ct severity scores in
  patients with suspected covid-19 using artificial intelligence.
\newblock {\em Radiology}, 2020.

\bibitem{Li20}
L.~Li, L.~Qin, Z.~Xu, Y.~Yin, X.~Wang, B.~Kong, J.~Bai, Y.~Lu, Z.~Fang,
  Q.~Song, et~al.
\newblock Artificial intelligence distinguishes {COVID}-19 from community
  acquired pneumonia on chest {CT}.
\newblock {\em Radiology}, 2020.

\bibitem{Li20a}
Y.~{Li}, W.~{Dong}, J.~{Chen}, S.~{Cao}, H.~{Zhou}, Y.~{Zhu}, J.~{Wu},
  L.~{Lan}, W.~{Sun}, T.~{Qian}, K.~{Ma}, H.~{Xu}, and Y.~{Zheng}.
\newblock Efficient and effective training of {COVID-19} classification
  networks with self-supervised dual-track learning to rank.
\newblock {\em IEEE Journal of Biomedical and Health Informatics}, 2020.

\bibitem{Ning20}
W.~Ning, S.~Lei, J.~Yang, Y.~Cao, P.~Jiang, Q.~Yang, J.~Zhang, X.~Wang,
  F.~Chen, Z.~Geng, et~al.
\newblock {iCTCF}: an integrative resource of chest computed tomography images
  and clinical features of patients with {COVID}-19 pneumonia.
\newblock {\em Research Square}, 2020.

\bibitem{Ouya20}
X.~Ouyang, J.~Huo, L.~Xia, F.~Shan, J.~Liu, Z.~Mo, F.~Yan, Z.~Ding, Q.~Yang,
  B.~Song, et~al.
\newblock Dual-sampling attention network for diagnosis of {COVID}-19 from
  community acquired pneumonia.
\newblock {\em IEEE Transactions on Medical Imaging}, 2020.

\bibitem{Prok20}
M.~Prokop, W.~van Everdingen, T.~van Rees~Vellinga, J.~Quarles~van Ufford,
  L.~Stoger, L.~Beenen, B.~Geurts, H.~Gietema, J.~Krdzalic, C.~Schaefer-Prokop,
  B.~van Ginneken, M.~Brink, and {COVID-19 Standardized Reporting Working Group
  of the Dutch Radiological Society}.
\newblock {CO-RADS} - a categorical {CT} assessment scheme for patients with
  suspected {COVID-19}: definition and evaluation.
\newblock {\em Radiology}, page 201473, 4 2020.

\bibitem{Ragh19a}
M.~Raghu, C.~Zhang, J.~Kleinberg, and S.~Bengio.
\newblock Transfusion: Understanding transfer learning for medical imaging.
\newblock In {\em Advances in Neural Information Processing Systems}, pages
  3342--3352, 2019.

\bibitem{Rutt00}
C.~M. Rutter.
\newblock Bootstrap estimation of diagnostic accuracy with patient-clustered
  data.
\newblock {\em Academic Radiology}, 7:413--419, 2000.

\bibitem{Samu07a}
F.~Samuelson, N.~Petrick, and S.~Paquerault.
\newblock Advantages and examples of resampling for {CAD} evaluation.
\newblock In {\em IEEE International Symposium on Biomedical Imaging}, pages
  492 --495, 2007.

\bibitem{Sing20a}
D.~Singh, V.~Kumar, and M.~Kaur.
\newblock Classification of {COVID}-19 patients from chest {CT} images using
  multi-objective differential evolution-based convolutional neural networks.
\newblock {\em European Journal of Clinical Microbiology \& Infectious
  Diseases}, pages 1--11, 2020.

\bibitem{Ying20}
Y.~Song, S.~Zheng, L.~Li, X.~Zhang, X.~Zhang, Z.~Huang, J.~Chen, H.~Zhao,
  Y.~Jie, R.~Wang, et~al.
\newblock Deep learning enables accurate diagnosis of novel coronavirus
  ({COVID-19}) with {CT} images.
\newblock {\em medRxiv}, 2020.

\bibitem{Sun20}
L.~{Sun}, Z.~{Mo}, F.~{Yan}, L.~{Xia}, F.~{Shan}, Z.~{Ding}, B.~{Song},
  W.~{Gao}, W.~{Shao}, F.~{Shi}, H.~{Yuan}, H.~{Jiang}, D.~{Wu}, Y.~{Wei},
  Y.~{Gao}, H.~{Sui}, D.~{Zhang}, and D.~{Shen}.
\newblock Adaptive feature selection guided deep forest for {COVID-19}
  classification with chest {CT}.
\newblock {\em IEEE Journal of Biomedical and Health Informatics}, 2020.

\bibitem{Ginn10a}
B.~van Ginneken, S.~G. Armato, B.~de~Hoop, S.~van~de Vorst, T.~Duindam,
  M.~Niemeijer, K.~Murphy, A.~M.~R. Schilham, A.~Retico, M.~E. Fantacci,
  N.~Camarlinghi, F.~Bagagli, I.~Gori, T.~Hara, H.~Fujita, G.~Gargano,
  R.~Belloti, F.~D. Carlo, R.~Megna, S.~Tangaro, L.~Bolanos, P.~Cerello, S.~C.
  Cheran, E.~L. Torres, and M.~Prokop.
\newblock Comparing and combining algorithms for computer-aided detection of
  pulmonary nodules in computed tomography scans: the {ANODE09} study.
\newblock {\em Medical Image Analysis}, 14:707--722, 12 2010.

\bibitem{Wang20b}
J.~Wang, Y.~Bao, Y.~Wen, H.~Lu, H.~Luo, Y.~Xiang, X.~Li, C.~Liu, and D.~Qian.
\newblock Prior-attention residual learning for more discriminative {COVID}-19
  screening in {CT} images.
\newblock {\em IEEE Transactions on Medical Imaging}, 2020.

\bibitem{Wang20a}
S.~Wang, B.~Kang, J.~Ma, X.~Zeng, M.~Xiao, J.~Guo, M.~Cai, J.~Yang, Y.~Li,
  X.~Meng, et~al.
\newblock A deep learning algorithm using {CT} images to screen for corona
  virus disease ({COVID}-19).
\newblock {\em medRxiv}, 2020.

\bibitem{Wang20}
S.~Wang, Y.~Zha, W.~Li, Q.~Wu, X.~Li, M.~Niu, M.~Wang, X.~Qiu, H.~Li, H.~Yu,
  et~al.
\newblock A fully automatic deep learning system for {COVID}-19 diagnostic and
  prognostic analysis.
\newblock {\em European Respiratory Journal}, 2020.

\bibitem{Xie20}
W.~Xie, C.~Jacobs, J.-P. Charbonnier, and B.~van Ginneken.
\newblock Relational modeling for robust and efficient pulmonary lobe
  segmentation in {CT} scans.
\newblock {\em IEEE Transactions on Medical Imaging}, 39:2664--2675, 2020.

\bibitem{Yang2020}
W.~Yang, A.~Sirajuddin, X.~Zhang, G.~Liu, Z.~Teng, S.~Zhao, and M.~Lu.
\newblock The role of imaging in 2019 novel coronavirus pneumonia ({COVID}-19).
\newblock {\em European Radiology}, pages 1--9, 2020.

\bibitem{Zhen20a}
C.~Zheng, X.~Deng, Q.~Fu, Q.~Zhou, J.~Feng, H.~Ma, W.~Liu, and X.~Wang.
\newblock Deep learning-based detection for {COVID}-19 from chest {CT} using
  weak label.
\newblock {\em medRxiv}, 2020.

\end{thebibliography}

\end{document}